\documentclass[intlimits,twoside,a4paper]{article}

\usepackage[cp1251]{inputenc}

\usepackage[eqsecnum]{cmpj3}

\issue{2024}{27}{3}{33802}
\doinumber{10.5488/CMP.27.33802}

\title[The comparison of female representation across mythologies]%
{Female representation across mythologies}
\author[M. Janickyj, P. MacCarron, J. Yose, \framebox{R. Kenna}]
{M. Janickyj \orcid{0009-0005-6113-3610}\refaddr{label1,label2,label4}\thanks{Corresponding author: \email{janickym@uni.coventry.ac.uk}.}, 
P. MacCarron \orcid{0000-0002-5163-9264}\refaddr{label3},
J. Yose \orcid{0000-0002-8250-213X} \refaddr{label1,label2},
\framebox{R. Kenna} \orcid{0000-0001-9990-4277} \refaddr{label1,label2}}

\addresses{
\addr{label1} Centre for Fluid and Complex Systems, Coventry University, Coventry, CV1 5FB, United Kingdom
\addr{label2} $\mathbb{L}^4$ Collaboration \& Doctoral College for the Statistical Physics of Complex Systems, Leipzig-Lorraine-Lviv-Coventry, Europe
\addr{label3} MACSI, Department of Mathematics and Statistics, University of Limerick, Limerick V94 T9PX, Ireland
\addr{label4} Department of Computer Science, University College London, London, UK
}


\date{Received January 15, 2024, in final form April 06, 2024}

\begin{document}

\maketitle

\begin{abstract}

Social groups have been studied throughout history to understand how different configurations impact those within them. Along with this came the interest in investigating social groups of both fictional and mythological works. Over the last decade these social groups have been studied through the lens of network science allowing for a new level of comparison between these stories. We use this approach to focus on the attributes of the characters within these networks, specifically looking at their gender. With this we review how the female populations within various narratives and to some extent the societies they are based in are portrayed. Through this we find that although there is not a trend of all narratives of the same origin having similar levels of representation some are noticeably better than others.  We also observe which narratives overall prioritise important female characters and which do not.

\printkeywords

\end{abstract}

\section{Introduction}

Complex networks have been used increasingly in recent years to represent various structures and systems. Some of these are physical networks such as those seen in transport systems, while others are social networks that depict the relationships between individuals. These complex networks allow for the quantitative analysis and comparison of various systems to be done which was first applied to mythological narratives in 2012~\cite{mac2012universal}. This method of creating social networks of myths, or narrative networks, describes the interactions between characters in these stories to give a quantitative approach to this piece of history.

One attribute of the characters that we often identify from the narratives is their gender. This can be used to evaluate how gender is represented across the whole network, and thus the narrative. This in turn can be used to further the comparisons between narrative networks and in some way the different societies and cultures, as depicted in the texts. 

In our current society, it is becoming more common to see a woman in some form of position of power, or being the focal point, whether this be within the real world or in some fictional medium. Along with this comes the larger, or more public, realisation that many of the stories read and studied so often feature males more prominently. Although none of the texts we have chosen to include focus solely on women, which is to be expected given the backgrounds/time frames of them all, we choose to investigate these texts to build upon the existing analysis and reveal any insight relating to how women are portrayed quantitatively within them. 

Many epic and mythological narratives come from patriarchal cultures. For example, in the sagas of the Icelanders and the \emph{Iliad}, it is very common for characters to identify themselves by their father's name. This is generally true of the Irish narratives as well where ``Mac'' in a character's name translates to ``son of''. However, some prominent characters there identify themselves from their mother's name --- the king of Ulster Conchobur Mac Nessa and one of the main gods Lugh Mac Eithne being major characters who do this.
Of the fictional narratives, \emph{A Song of Ice and Fire} is intentionally constructed as a patriarchal society. Therefore, we might expect the Irish narratives to have more prominent female characters from a network perspective than some of the other datasets. 

In this work we compare 21 narratives from various cultures, 2 of them being fictional, with the intention of determining how women are represented in the different stories and to some extent in the societies presented in them. Some of these narratives have previously been examined but we include these data to gain a better understanding of the different levels of representation. The data of these texts have been collected by the authors and have been discussed, to some degree in \cite{mac2012universal, mac2013network, mac2014network, maccarronPhD2014, yosePhD2017, yose2016ossian, gessey2020narrative, epic2023}.
However, little work has been done on the gender of the characters in these. A similar approach to this is done in \cite{prado2020gendered}, where they are interested in determining whether females in the network reduce the male violence. 
There they want to know how many shortest paths the female nodes lie on, which is similar to the concept of betweenness centrality that we use here.
We also see similar approaches in \cite{mattison2021gender}, where betweenness centrality, and other measures, are used to evaluate the social relationships of genders across networks. Lastly, as mentioned above, some of these networks had been investigated previously to different degrees. In \cite{epic2023} some had been compared to judge the most influential characters of the network, focusing on their gender, and how this impacts the narrative. This approach is similar to the one we will be taking.  

The paper is structured as follows: in the next section we describe the methods used and give an overview of the data, then within the results section we use attribute-based network measures to investigate whether the networks represent female characters well. Within the conclusion we evaluate the trends found across the narratives and the work as a whole.

\section{Methods}

To properly compare these narratives in a quantitative sense, the characters and their interactions are identified through a thorough reading of the texts. This follows the methodologies observed in the references mentioned above, where characters are determined to be connected if they meet each other or if a previous interaction is made clear within the text. From this collection of data, characters and edges are then treated as nodes and edges respectively to construct these narrative networks

Along with the social interactions between characters, further information is extracted that can be used to provide additional insight into the characters and their society. In our case, this is the gender, though another example is the polarity of an interaction. Specifically, we classify interactions as being friendly or hostile. A hostile relation indicates that there is some sort of violence between the characters, such as them fighting in a battle, or they are explicitly clear about their animosity. The latter could be observed when a character is threatened by another. Other pieces of information that are collected depend on the type of society that exists within the narrative. This could include the group or faction a character is from, which is helpful when examining the networks that focus on major battles, such as that of the \textit{Iliad} or of \textit{T\'{a}in B\'{o} C\'{u}ailnge}. {{Within this work, we consider the full network comprised of both friendly and hostile edges. Although it would be interesting to discuss this element of importance and gender representation alongside the hostile and friendly networks, we also acknowledge that historically men are more often the warriors \citep{prugl2003gender}. This is echoed by the fact that there are very few, if any, hostile female-female interactions throughout these narratives.}}

The attribute that we will be mainly focusing on, is the gender of the characters and how this is represented in the whole network, and thus the narrative. Investigating the representation of females within these narratives using networks adds another element of quantitative analysis to these epics and provides another layer of comparison between the societies and cultures written with them.

\subsection{Network Measures}

Here, we introduce some attribute specific network measures used to conduct this analysis. While we give a short overview in the results section regarding the global measure, the main focus is on the attribute specific, or gender specific results. To investigate the representation and importance of female characters within the network, we evaluate some measures relating to the specific nodes, while others focus on the edges of the network. In both cases, we look at how the females are connected to the network. A fundamental measure of a node is its number of connections, or its degree $k$. The degree $k$ of each node is in turn used to calculate the mean degree, $\langle k \rangle$, and is defined as $2L/N$ where $L$ and $N$ represent the number of edges and nodes in the network, respectively. The degree of an individual node will be compared with the average of the network to judge which nodes are well-connected overall.

Another measure of the influence of a node is its betweenness centrality. Betweenness centrality determines a node to be important if it connects other nodes along the shortest path and is useful to other nodes in the network. In a network, a path is a sequence of consecutive edges that connects some pair of nodes. The shortest path is the path between any pair of nodes containing the fewest number of edges. Using this measure, we are able to isolate the top $10\%$ of the network to see how females exist within this space.

In many forms of media, the Bechdel test is used to evaluate how female characters are portrayed~\cite{bechdel_1985}. This test includes three criteria; there must be two women in the piece of media, they must talk to each other, and the topic of conversation must be something other than a man. Although this is not a set of criteria that we can evaluate networks with, it is possible to apply these principles using the edges of a network. The proportion of edges that connect only female, male, or both characters will be used to determine how females interact within the networks.

\subsection{The chosen narratives}

The majority of the 21 chosen narratives compared within this study are mythological texts or epic literature, though the two are purely fiction. The latter are the \textit{The Lord of the Rings} (LotR) trilogy written by J. R. R. Tolkien \cite{LotR1995} and the first five books of \textit{A Song of Ice and Fire} series by George R. R. Martin~\cite{GoT1, GoT2, GoT3, GoT4, GoT5}. These are both pieces of fantasy literature with the first telling of a groups' journey to destroy a magical ring and the other of the ruling houses of the land Westeros. 

The included mythological narratives are from 10 different cultures or regions. The most common of which are those of Greek (three texts), Irish (four), or Icelandic (five) origins. The other seven narratives are all the only texts included from their culture. This includes narratives such as the \textit{Mahabharata} which is of Indian origins \cite{Mahabharata2009} and \textit{Popol Vuh}, a Central American myth \cite{PopolVuh1991}. The other narratives and their backgrounds will be introduced throughout this work. {{However, for ease, we list all 21 narratives, their origins, and some basic network measures in table \ref{Tab20origins}.}}

{{With many of the stories considered in this work being of various backgrounds, it is also worth discussing any major differences between the composition, or content, of the narratives. Of the 21 narratives, four are compilations of short stories rather than a singular narrative. However, almost all of these four works (the \textit{Mahabharata}, \textit{The Poems of Ossian}, \textit{Acallam an Sen\'{o}rach}, and the Ulster cycle) contain stories which tell us of events concerning many of the same characters, typically over a short period of time. This in turn means that while the way the stories are told differs in some regard, how the characters interact as a whole and how this translates into the network will differ in the same way the type of story impacts the social network. Additionally, some of the narratives here tell us of wars, meaning there may be a higher proportion of hostile interactions than is seen in other narrative types. However, while these networks (the \textit{Iliad}, \textit{Nj\'{a}l's Saga} and \textit{T\'{a}in B\'{o} C\'uailnge}), and those compiled of short stories, may contain different overall network properties, these differences are not seen when considering their gender representation.}}

\begin{table}[ht]
\begin{center}
\caption{List of 21 included narratives and their origins. Some names have been shortened for ease; \textit{Acallam an Sen\'{o}rach} is written as \textit{Acallam}, \textit{Lady Gregory} to \textit{Gods and Fighting Men: The Story of the Tuatha De Danaan and of the Fianna of Ireland, Arranged and put into English by Lady Augusta Gregory}, and \textit{Ulster} to an amalgamation of various texts from the Ulster cycle, though it does not include \textit{T\'{a}in B\'{o} C\'uailnge}. \\ Here, the numbers of nodes and edges are denoted by $N$ and $L$, respectively. The quantity $\langle k \rangle$ is the mean degree of the network and  $C$ is its mean clustering coefficient. The mean path length is given by $\ell$. The assortativity is denoted by $r$ and $G_{C}$ is the relative size of the giant component.}
	\vspace{2mm}
\begin{tabular}{|c|c|c|c|c|c|c|c|c|}
\hline
 Network & Origin & $N$ & $L$ & $\langle k \rangle$ & $C$ & $\ell$ & $r$ & $G_{C} (\%)$ \\ \hline
\textit{A Song of Ice and Fire} & Fiction & 1786 & 14478 & 16.21 & 0.65 & 3.33 & $-0.03$ & 94.63 \\ \hline
\textit{Acallam} & Irish & 806 & 1614 & 4.01 & 0.34 & 3.79 & $-0.10$ & 69.85 \\ \hline
\textit{Aeneid} & Greek & 538 & 985 & 3.66 & 0.34 & 3.57 & $-0.13$ & 72.86 \\ \hline
\textit{Arthurian Romances} & Welsh & 258 & 714 & 5.54 & 0.66 & 2.92 & $-0.29$ & 98.06 \\ \hline
\textit{Beowulf} & Anglo-Saxon & 74 & 167 & 4.51 & 0.56 & 2.38 & $-0.12$ & 67.57 \\ \hline
\textit{Egils Saga} & Icelandic & 294 & 770 & 5.24 & 0.56 & 4.19 & $-0.07$ & 96.60 \\ \hline
\textit{The Epic of Gilgamesh} & Sumerian & 56 & 81 & 2.89 & 0.38 & 2.54 & $-0.34$ & 76.79 \\ \hline
\textit{G\'{i}sli S\'{u}rsson's Saga} & Icelandic & 104 & 254 & 4.89 & 0.59 & 3.39 & $-0.15$ & 97.12 \\ \hline
\textit{Iliad} & Greek & 697 & 2686 & 7.71 & 0.44 & 3.49 & $-0.08$ & 99.00 \\ \hline
\textit{Lady Gregory} & Irish & 341 & 772 & 4.53 & 0.40 & 3.27 & $-0.17$ & 87.98 \\ \hline
\textit{Laxd{\ae}la Saga} & Icelandic & 337 & 894 & 5.31 & 0.44 & 5.01 & 0.19 & 97.63 \\ \hline
\textit{Lord of the Rings} & Fiction & 246 & 892 & 7.25 & 0.57 & 2.90 & $-0.20$ & 85.77 \\ \hline
\textit{Mahabharata} & Indian & 537 & 2666 & 9.93 & 0.51 & 3.56 & -0.05 & 96.65 \\ \hline
\textit{Navaho Indian Myths} & Navaho & 148 & 285 & 3.85 & 0.42 & 3.86 & $-0.18$ & 89.19 \\ \hline
\textit{Nj\'{a}l's Saga} & Icelandic & 577 & 1612 & 5.59 & 0.42 & 5.14 & 0.01 & 99.65 \\ \hline
\textit{Odyssey} & Greek & 307 & 1020 & 6.65 & 0.44 & 3.29 & $-0.08$ & 96.42 \\ \hline
\textit{The Poems of Ossian} & Scottish & 340 & 745 & 4.38 & 0.41 & 3.84 & -0.08 & 88.24 \\ \hline
\textit{Popol Vuh} & Mayan & 103 & 409 & 7.94 & 0.52 & 2.80 & $-0.32$ & 88.24 \\ \hline
\textit{T\'{a}in B\'{o} C\'uailnge} & Irish & 419 & 1263 & 6.03 & 0.73 & 2.78 & -0.35 & 98.33 \\ \hline
\textit{Ulster} & Irish & 234 & 1070 & 9.15 & 0.59 & 3.02 & $-0.03$ & 96.15 \\ \hline
\textit{Vatnsd{\ae}la's Saga} & Icelandic & 132 & 290 & 4.39 & 0.49 & 3.86 & 0.00 & 96.97 \\ 
\hline
\end{tabular}
\label{Tab20origins}
\end{center}
\end{table}

\section{Results}

\subsection{Standard network measures}

Although the focus of this work is on the representation of gender within these networks and the narratives they depict, it is interesting to first review the global network measures seen within table \ref{Tab20origins}. We first see the range in size of networks (meaning the number of nodes present within them). Spanning across five novels \textit{A Song of Ice and Fire} is the largest network studied here and contains $1786$ characters. Focusing on the other extreme shows \textit{The Epic of Gilgamesh} to be the smallest network with only 56 characters included. 

Some of the other measures shown in table \ref{Tab20origins}, such as the average degree, depend upon the size of the network. With this measure we can see how connected the nodes of these networks are and whether there are any similarities between them. Initially focusing on those narratives that are of the same background (those which are Greek, Irish, or Icelandic) indicates that of the three origins, the Icelandic narratives are those to have the closest average degree across all narratives. Though this doesn't tell us of the similarities between the stories themselves (and is dependent on the number of nodes and edges in the network), it is interesting.

Another way of judging how well connected the overall networks are is done by using the Giant Component of the network (the largest connected component of the network). In table \ref{Tab20origins} this is indicated by the percentage of nodes that are present in the Giant Component. Many of the networks investigated here are seen to have Giant Components of over $90\%$, the largest being that of \textit{Nj\'{a}l's Saga} which contains $99.65\%$ of its characters. The focus of this narrative is a blood feud between two families and so it is not unexpected that the majority of the characters have interacted with each other indirectly throughout the saga. Similarly, it is not unexpected that \textit{Beowulf} has the smallest giant component of $67.57\%$. This is a narrative that deals with both a main story but includes two smaller stories addressing past events \cite{mac2012universal}.

With a range of $0.40$ between the most and least clustered networks, the average clustering coefficient is one measure where we do not see many anomalous results. This is an indication of how the nodes form groups within the networks. However, even with this relatively low range across the 21 narratives, the five of those of Icelandic background still possess more similar values than those of other similar origins, namely the Greek and Irish epics. Focusing now on the attribute-specific measures may indicate whether these similarities are also exhibited there.

\subsection{Gender representation in the networks}

An initial judgement of the gender representation within these datasets is performed using the overall gender split of the network. As seen in table~\ref{Tab20gender}, there are characters within some of the networks whose genders are not known. Only three of the narratives in table \ref{Tab20gender} contain none of these characters. These narratives are \textit{Beowulf} \cite{BeowulfH1999}, \textit{Nj\'{a}l's Saga} \cite{bayerschmidt1998njal} and \textit{Lady Gregory}'s text \cite{LadyGregory}. \textit{Beowulf}, an Anlgo-Saxon text, tells us of the hero Beowulf who comes to help the Danish king Hrothgar battle the monster known as Grendel. \textit{Nj\'{a}l's Saga} is one of 18 Icelandic Sagas and contains the story of a blood feud between two families which spans over 50 years. \textit{Lady Gregory} is an Irish text which consists of parts of the Fenian cycle and was arranged by Lady Gregory.

Although the other 18 narratives include characters where the gender is unknown, the percentage of these types of characters does not exceed $25\%$ for the majority of the texts ($33.01\%$ of \textit{Popol Vuh}'s nodes are of unknown gender).

\begin{table}[ht]
\begin{center}
\caption{Gender-specific properties of multiple narrative networks, where the $\%f, \%u$ and $\%m$ represent the percentage of female, unknown and male nodes, respectively. The percentage of female nodes who have a degree greater than the network's average is denoted by `$\%N_{f} k> \langle k \rangle$', and `$\%N_{m} k> \langle k \rangle$' describes the percentage of male characters that fit this requirement, '$N_{f}:N_{m}$' indicates the ratio of female to male nodes in the networks.}
\vspace{2mm}
\resizebox{\columnwidth}{!}{
\begin{tabular}{|c|c|c|c|c|c|c|}
\hline
 Network & $\%f$ & $\%u$ & $\%m$ & $\%N_{f} k> \langle k \rangle$ & $\%N_{m} k> \langle k \rangle$ & $N_{f}:N_{m}$\\ \hline
\textit{A Song of Ice and Fire} & 16.46 & 14.67 & 68.87 & 30.27 & 33.09 & 1:4.18 \\ \hline
\textit{Acallam} & 11.66 & 3.60 & 84.74 & 29.79 & 23.87 & 1:7.27 \\ \hline
\textit{Aeneid} & 18.59 & 2.60 & 78.81 & 27.00 & 30.90 & 1:4.24 \\ \hline
\textit{Arthurian Romances} & 19.77 & 1.55 & 78.68 & 23.53 & 21.67 & 1:3.98 \\ \hline 
\textit{Beowulf} & 12.16 & 0.00 & 87.84 & 66.67 & 38.46 & 1:7.22 \\ \hline
\textit{Egils Saga} & 16.67 & 1.36 & 81.97 & 38.78 & 29.88 & 1:4.92 \\ \hline
\textit{The Epic of Gilgamesh} & 23.21 & 8.93 & 67.86 & 38.46 & 31.58 & 1:2.92 \\ \hline
\textit{G\'{i}sli S\'{u}rsson's Saga} & 16.35 & 0.96 & 82.69 & 70.59 & 30.23 & 1:5.06 \\ \hline
\textit{Iliad} & 13.92 & 18.65 & 67.43 & 44.33 & 22.55 & 1:4.85 \\ \hline
\textit{Lady Gregory} & 13.49 & 0.00 & 86.51 & 32.61 & 29.49 & 1:6.41 \\ \hline 
\textit{Laxd{\ae}la Saga} & 21.96 & 1.19 & 76.85 & 47.30 & 24.71 & 1:3.50 \\ \hline
\textit{Lord of the Rings} & 8.13 & 9.76 & 82.11 & 15.00 & 31.19 & 1:10.10 \\ \hline
\textit{Mahabharata} & 16.57 & 0.93 & 82.50 & 12.36 & 27.09 & 1:4.98 \\ \hline 
\textit{Navaho Indian Myths} & 27.03 & 21.62 & 51.35 & 42.50 & 35.53 & 1:1.90 \\ \hline
\textit{Nj\'{a}l's Saga} & 16.46 & 0.00 & 83.54 & 21.05 & 24.48 & 1:5.07 \\ \hline
\textit{Odyssey} & 25.08 & 1.63 & 73.29 & 28.57 & 31.56 & 1:2.92 \\ \hline
\textit{The Poems of Ossian} & 16.77 & 2.06 & 81.18 & 17.54 & 28.99 & 1:4.84 \\ \hline
\textit{Popol Vuh} & 11.65 & 33.01 & 55.34 & 33.33 & 57.89 & 1:4.75 \\ \hline
\textit{T\'{a}in B\'{o} C\'uailnge} & 8.59 & 20.53 & 70.88 & 25.00 & 25.59 & 1:8.25 \\ \hline
\textit{Ulster} & 17.52 & 5.98 & 76.50 & 29.27 & 24.02 & 1:4.37 \\ \hline
\textit{Vatnsd{\ae}la's Saga} & 21.21 & 0.76 & 78.03 & 21.43 & 31.07 & 1:3.68 \\
\hline
\end{tabular}
}
\label{Tab20gender}
\end{center}
\end{table}

Focusing first on the percentage of female nodes within the network (overall nodes in the network and including those of unknown gender), it can be seen that the narratives with the top three percentages, the \textit{Navaho Indian Myths} \cite{Navaho1993}, the \textit{Odyssey} \cite{Odyssey1980} and \textit{The Epic of Gilgamesh} \cite{Gilgamesh2002}. $27.03\%$, $25.08\%$ and $23.21\%$ of these networks' nodes respectively are female. Even when only considering the nodes of identifiable gender, these networks have the three highest percentages of women and these networks are of different origins; \textit{Navaho Indian Myths} from the Native American tribe of the same name, the \textit{Odyssey} is a Greek epic and \textit{The Epic of Gilgamesh} is an early Sumerian text. The \textit{Navaho Indian Myths} are a collection of tales told by a chief and contain various important stories including the creation. The \textit{Odyssey} is one of Homer's epics which tells of Odysseus, the King of Ithica, on his journey home after the Trojan War and the \textit{Epic of Gilgamesh} tells of Gilgamesh's journey to find the secret of eternal life.

Conversely, the three narratives with the smallest percentage of women overall are the \textit{LotR} trilogy, \textit{T\'{a}in B\'{o} C\'uailnge (TBC)}, and \textit{Popol Vuh}. $8.13\%$, $8.59\%$ and $11.65\%$ of these three networks are represented by female nodes, respectively. Like the three networks with the highest percentages of women, these are all set in different cultures and societies. One of which is even set in the fictional Middle Earth. \textit{TBC} refers to one of the most famous stories found within Irish mythology and along with its pretales tells us of Ulster being attacked by Queen Medb but being defended by C\'{u}chulainn. Like the \textit{Navaho Indian Myths}, \textit{Popol Vuh} is a collection of stories, and it features their story of the creation. The differing origins between these three narratives do not tell us initially whether one specific culture has a better (or worse) representation of gender within their narratives since both the top (and bottom) three are all of different backgrounds. However, comparing important nodes in the network provides more insight.

\subsection{Well connected women} \label{wellconnected}

Table \ref{Tab20gender} displays the percentage of male and female nodes that are well-connected due to their degree being larger than the networks' average. {We choose the mean degree here because \cite{maccarronPhD2014} identifies that many of these datasets are well fitted by exponential, Weibull, or log-normal distributions, all of which have a parameter related to the mean degree. It is also worth noting, that although all of these distributions are right-skewed and the median is often more representative in skewed data, the median is smaller in almost all of these datasets than the mean degree and we wish to use smaller sets of importance where possible.
}

The narrative with the highest percentage of well-connected female characters is the Icelandic \textit{G\'{i}sli S\'{u}rsson's Saga} \cite{IcelandS2000} of which $70.59\%$ of females have a higher than average degree. This narrative focuses on the central character G\'{i}sli who is being hunted and is on the run for thirteen years. $66.67\%$ of female characters in \textit{Beowulf} are more connected than the average nodes, and the network with the third highest percentage of well-connected female characters contains just under $20\%$ less females. $47.30\%$ of the female characters in \textit{Laxd{\ae}la} have an higher than average degree. This saga is another Icelandic narrative and tells us of people in the west of Iceland over the course of a century \cite{IcelandS2000}. Table \ref{Tab20gender} also shows that the network with the lowest percentage of important female characters is the \textit{Mahabharata} (12.36\%). This text is an ancient Indian narrative which tells us of two sides of a family and their battle with each other to rule.

The network with the highest proportion of well-connected men is the Central American narrative \textit{Popol Vuh} with $57.89\%$ of its male characters having a higher than average degree. The network with the next highest percentage, $38.46\%$, is \textit{Beowulf} which also had a high percentage of well-connected female characters. This is $2.94\%$ higher than the next narrative, \textit{Navaho Indian Myths}. While this narrative did not rank highly in the same measure with regard to female characters, it was still in the top half of the narratives. The network with the lowest percentage of important men is the \textit{Arthurian Romances}. The \textit{Arthurian Romances} are a collection of five Welsh narratives that each centres on a different knight \cite{Arthurian1991}. In this network, $21.67\%$ of men have a degree higher than the average which is still higher than the four networks with the lowest percentage of important females.

As displayed in table \ref{Tab20origins}, the network of the \textit{Mahabharata} and the \textit{Arthurian Romances} have average degrees of $9.93$ and $5.54$ respectively. The average degree of \textit{Mahabharata} is the second highest of all the networks discussed here. The highest is that of \textit{A Song of Ice and Fire} (16.21). Of the $537$ nodes in \textit{Mahabharata}, $24.39\%$ of them have a greater than average degree and the majority of these are male. Similarly, $21.71\%$ of the characters in the \textit{Arthurian Romances} have a greater than average degree but the split between males and females in these well-connected characters is less drastic than seen in \textit{Mahabharata} (1 woman to 3.7 men compared with 1 woman to 10.9 men). This shows that, at least for the \textit{Arthurian Romances}, it is not necessarily just women that are not well connected.

Comparison of the overall percentages of well-connected characters (without considering their genders) shows a mixture of narratives that had well-connected males and females to be present in these top three. $41.89\%$ of the characters in \textit{Beowulf} have a degree greater than $4.51$ which is the network's average. Of \textit{G\'{i}sli S\'{u}rsson's Saga}'s 105 nodes, $36.54\%$ are well-connected. Though over $70\%$ of the women of this network were well-connected only $30.23\%$ of its men are. \textit{Popol Vuh} has the third highest percentage of well-connected nodes overall (35.92$\%$) and also had the highest proportion of men with a greater than average degree. By discussing the proportion of nodes that are well connected in the network, we have investigated whether females are seen to be important in this regard.

\subsection{Identifying important characters using betweenness centrality}

Table \ref{Tab20BC} shows both the percentage split of the top $10\%$ of the network and the percentage of these nodes out of the total number of their respective gender. This, like the measure of well-connected characters, displays the percentage of the characters of that gender that are important. The gender split of the top $10\%$ of nodes can also be used to judge whether it is reflective of the gender split of the whole network. One initial difference between the overall gender split and split of the important nodes is that the majority of the networks include nodes of unknown gender while all of the important characters are either male or female. \textit{Beowulf} and \textit{The Poems of Ossian} are the only narratives to contain no women in their top $10\%$ of nodes, meaning by this measure that none of its $9$ and $57$ female characters, respectively, are important. However, a high percentage of \textit{Beowulf}'s female nodes are well-connected, as seen previously. The same cannot be said for the female characters of \textit{The Poems of Ossian}, which is referred to as \textit{Ossian} within this work \cite{Ossian1762, Ossian1763, Ossian1765, Ossian1773}.
 
The three narratives that contain the highest percentage of female characters within the most important $10\%$ are the \textit{Odyssey}, the \textit{Navaho Indian Myths} and the \textit{Arthurian Romances}. The \textit{Navaho Indian Myths} and the \textit{Odyssey} were also in the top three networks for the percentage of women overall. The \textit{Arthurian Romances} was the network with the 6th highest percentage of women. Comparing these statistics with the percentage of women in the networks that are important, indicates whether this is representative of the women in the full network or whether there just happen to be a few women that are important while many others are not.

\begin{table}[ht]
\begin{center}
\caption{Gender-specific properties of multiple narrative networks focussing on measures relating to betweenness centrality. Here, `$10\%$` is the number of nodes that make up the top $10\%$ of the network when ranked by their betweenness centrality, `$\%_{f}$ in top $10\%$' and '$\%_{m}$ in top $10\%$' are the percentage of the top 10\% which are female and male respectively, `$\%$ of important women' and `$\%$ of important men' represent the percentage of men and women that are deemed important because they are in the top $10\%$ of the network due to their betweenness centrality. {Values in bold represent where the percentage of important females is higher than males.}}
\vspace{2mm}
\resizebox{\columnwidth}{!}{
\begin{tabular}{|c|c|c|c|c|c|c|c|}
\hline
 Network & $10\%$ & $\%_{f}$ in top $10\%$ & $\%_{m}$ in top $10\%$ & $\%$ of important women & $\%$ of important men \\ \hline
\textit{A Song of Ice and Fire} & 174 & 11.49 & 88.51 & 6.80 & 12.52 \\ \hline
\textit{Acallam} & 79 & 16.46 & 83.54 & \textbf{13.83} & 9.66 \\ \hline
\textit{Aeneid} & 55 & 14.55 & 85.45 & 8.00 & 11.08 \\ \hline
\textit{Arthurian Romances} & 27 & 25.93 & 74.07 & \textbf{13.73} & 9.85 \\ \hline 
\textit{Beowulf} & 9 & 0.00 & 100.00 & 0.00 & 13.85 \\ \hline
\textit{Egils Saga} & 31 & 16.13 & 83.87 & 10.20 & 10.79 \\ \hline
\textit{The Epic of Gilgamesh} & 7 & 14.29 & 85.71 & 7.69 & 15.79 \\ \hline
\textit{G\'{i}sli S\'{u}rsson's Saga} & 12 & 8.33 & 91.67 & 5.88 & 12.79 \\ \hline
\textit{Iliad} & 69 & 11.59 & 88.41 & 8.25 & 12.98 \\ \hline
\textit{Lady Gregory}  & 36 & 16.67 & 83.33 & \textbf{13.04} & 10.17 \\ \hline 
\textit{Laxd{\ae}la Saga} & 35 & 20.00 & 80.00 & 9.46 & 10.81 \\ \hline
\textit{Lord of the Rings} & 25 & 4.00 & 96.00 & 5.00 & 11.88 \\ \hline
\textit{Mahabharata} & 54 & 5.56 & 94.44 & 3.37 & 11.51 \\ \hline 
\textit{Navaho Indian Myths} & 16 & 31.25 & 68.75 & 12.50 & 14.47 \\ \hline
\textit{Nj\'{a}l's Saga} & 59 & 10.17 & 89.83 & 6.32 & 11.00 \\ \hline
\textit{Odyssey} & 32 & 34.38 & 65.63 & \textbf{14.29} & 9.33 \\ \hline
\textit{The Poems of Ossian} & 35 & 0.00 & 100.00 & 0.00 & 12.68 \\ \hline
\textit{Popol Vuh} & 11 & 9.09 & 90.91 & 8.33 & 17.54 \\ \hline
\textit{T\'{a}in B\'{o} C\'uailnge} & 42 & 16.67 & 83.33 & \textbf{19.44} & 11.78 \\ \hline
\textit{Ulster} & 24 & 8.33 & 91.67 & 4.88 & 12.29 \\ \hline
\textit{Vatnsd{\ae}la's Saga} & 15 & 13.33 & 86.67 & 7.14 & 12.62 \\
\hline
\end{tabular}
}
\label{Tab20BC}
\end{center}
\end{table}

\textit{TBC} is the network with the highest percentage of important women. Of the 36 women in \textit{TBC}, seven are ranked in the top $10\%$ of the network by their betweenness centrality which equates to $19.44\%$. For this network, $16.67\%$ of the nodes in the top $10\%$ of the whole network are female. Interestingly, \textit{TBC} has one of the lowest percentages of women out of all the networks in table \ref{Tab20gender}. This does tell us that even with few women in the network overall, a higher proportion of them are important to the story and to others in the network. We also see that a smaller percentage of men than women are determined to be important in \textit{TBC}. This is partly because \textit{TBC} is quite patriarchal in that characters usually identify themselves as "son of...", therefore we get a lot of low degree (and so typically of low betweenness centrality) characters that are not present in the story, but are named fathers, thus male.

The next network, when ranked by the percentage of important women, is the \textit{Odyssey} where $14.29\%$ of women are deemed important. The \textit{Odyssey} is one of the top three networks when ranked on percentage of women overall, well-connected women, and women in the top $10\%$ of the network ($34.38\%$ of the important characters are women). This shows that it not only has a good percentage of women in relation to the other networks but also shows, that a higher percentage of women in this network are important too. 

The network that is ranked third with respect to important women is \textit{Acallam an Sen\'{o}rach}, this is an Irish narrative that tells us of two heroes that recount their travels to Saint Patrick \cite{DooleyRow1999}. The stories they tell often relate to a rivalry between their family and their enemies. This narrative will be known as \textit{Acallam}. $13.83\%$ of the 94 women in this network are deemed important. Like \textit{TBC}, this network contains a smaller percentage of women when compared with the other networks but, unlike \textit{TBC} only $3.60\%$ of this network's characters' genders are undefined. This tells us that for these two networks, though there are not many women present in the narratives overall, the female characters that are in the network are important. For example, one of the main characters in \textit{TBC} is Queen Medb. 

\subsection{Interactions involving female characters}

A measure of female presence within a narrative can also be judged using the percentage of interactions that involve at least one female character. The network with the highest proportion of edges to connect a woman to some other node is the network of the \textit{Navaho Indian Myths}. Of this network's $285$ edges, $55.44\%$ link two nodes where at least one is a woman. The two networks with the next highest percentage are the \textit{Laxd{\ae}la Saga} \cite{IcelandS2000} and the \textit{Arthurian Romances}.
$44.52\%$ and $37.82\%$ of these networks' edges respectively involve a woman.
\textit{Laxd{\ae}la Saga} is also ranked within the top three networks when considering the percentage of well-connected women. This indicates that, as mentioned in section \ref{wellconnected}, it seems to have a good level of female representation.

\begin{table}[h]
\begin{center}
\caption{Gender-specific properties of multiple narrative networks concerning different edge types where $L_{f-\textrm{any}}$, $L_{f-f}$, $L_{m-m}$ and $L_{m-f}$ indicate the percentage of edges involving at least one female character, two females, two males and a male and female, respectively. The two columns $L_{f-u}$ and $L_{m-u}$ indicate the percentage of edges involving a character of unknown gender and a female and a male, respectively. {We use bold when more than one third of the interactions involve a female character.}
}
\vspace{2mm}
\resizebox{0.7\textwidth}{0.9\height}{
\begin{tabular}{|c|c|c|c|c|c|c|}
\hline
 Network & $L_{f-any}$ & $L_{f-f}$ & $L_{m-m}$ & $L_{m-f}$ & $L_{f-u}$ & $L_{m-u}$ \\ \hline
\textit{A Song of Ice and Fire} & 29.98 & 4.93 & 56.90 & 21.99 & 3.07 & 11.17 \\ \hline
\textit{Acallam} & 20.69 & 1.49 & 73.23 & 18.71 & 0.50 & 5.45 \\ \hline
\textit{Aeneid} & 25.38 & 5.99 & 74.11 & 19.19 & 0.20 & 0.51 \\ \hline
\textit{Arthurian Romances} & \textbf{37.82} & 4.20 & 61.77 & 33.33 & 0.28 & 0.42 \\ \hline
\textit{Beowulf} & 27.55 & 1.20 & 72.46 & 26.35 & 0.00 & 0.00 \\ \hline
\textit{Egils Saga} &\textbf{ 33.38 }& 3.77 & 65.71 & 29.61 & 0.00 & 0.78 \\ \hline
\textit{The Epic of Gilgamesh} & \textbf{34.57} & 2.47 & 61.73 & 29.63 & 0.00 & 3.70 \\ \hline
\textit{G\'{i}sli S\'{u}rsson's Saga} &\textbf{ 33.47} & 3.15 & 66.14 & 30.32 & 0.00 & 0.39 \\ \hline
\textit{Iliad} & \textbf{34.36} & 20.29 & 54.58 & 10.95 & 3.13 & 9.94 \\ \hline
\textit{Lady Gregory} & 20.60 & 1.68 & 79.40 & 18.91 & 0.00 & 0.00 \\ \hline
\textit{Laxd{\ae}la Saga} &\textbf{ 44.52 }& 7.94 & 55.15 & 36.13 & 0.45 & 0.34 \\ \hline
\textit{Lord of the Rings} & 10.65 & 0.34 & 77.35 & 9.75 & 0.56 & 10.65 \\ \hline
\textit{Mahabharata} & 18.79 & 1.65 & 80.68 & 17.10 & 0.04 & 0.53 \\ \hline
\textit{Navaho Indian Myths} & \textbf{55.44} & 9.47 & 29.83 & 40.00 & 5.96 & 8.42  \\ \hline
\textit{Nj\'{a}l's Saga} & 26.68 & 3.10 & 73.33 & 23.57 & 0.00 & 0.00 \\ \hline
\textit{Odyssey} & 32.84 & 7.94 & 66.86 & 24.71 & 0.20 & 0.20 \\ \hline
\textit{The Poems of Ossian} & 22.82 & 0.67 & 73.56 & 21.75 & 0.40 & 3.62 \\ \hline
\textit{Popol Vuh} & 21.52 & 0.49 & 45.23 & 18.58 & 2.44 & 31.78 \\ \hline
\textit{T\'{a}in B\'{o} C\'uailnge} & 20.59 & 2.38 & 60.33 & 17.34 & 0.87 & 16.55 \\ \hline
\textit{Ulster} & 20.47 & 5.61 & 76.45 & 14.11 & 0.75 & 2.71 \\ \hline
\textit{Vatnsd{\ae}la's Saga} &\textbf{ 35.17} & 2.41 & 64.48 & 32.76 & 0.00 & 0.34 \\
\hline
\end{tabular}
}
\label{Tab20genderedges}
\end{center}
\end{table}

The narrative constructed from the \textit{LotR} trilogy has the smallest percentage of edges that involve women (10.65\%). It is also the network with the lowest percentage of female characters overall ($8.13\%$). This is not an unexpected occurrence since \textit{LotR} is a trilogy that follows a group of male characters on their journey to destroy the one ring. On this journey, there are certain female characters they meet that are important but these characters only make up $4.00\%$ of the top $10\%$ of the network's nodes. This $4.00\%$ represents just one female which is the elf Galadriel. Galadriel wears one of the rings of power and rules her realm (Lothl\'{o}rien) with her husband. The networks of the \textit{Mahabharata} and \textit{Popol Vuh} are the networks with the second and third lowest proportion of edges involving women. Only $18.79\%$ and $21.52\%$ of these networks' edges respectively fall into this category. The networks also do not show good representation of women in other areas of analysis. For example, only $3.37\%$ of women in \textit{Mahabharata} are deemed important.

{{In table \ref{Tab20genderedges} we also observe the percentage of edges that connect a male or female character with that of an unknown gender character. It is not shocking to observe that for the majority of narratives more edges involving a node of unknown gender link with a man rather than a woman. The only narrative that does not show this is \textit{Laxd{\ae}la Saga}, again showing it to have a good level of female representation. 

When focusing solely on the interactions that occur between 2 female characters we see that this includes $20.29\%$ of the \textit{Iliad}'s connections, a much higher figure than other networks. While this seems notable, it is also important to understand the context behind this figure. Within the \textit{Iliad}, there exists a clique of 33 \emph{Neirids} (female sea nymphs), of which Achilles' mother Thetis is one. The majority of these Neirids (barring Thetis) only interact with each other and as such, from this collection of characters, we gain around $18\%$ of the network's connections.}}

\section{Representation of women across cultures or societies}

The depiction of women in the networks studied here indicated that there was not an overall trend of one society or culture representing women better than others. The network in which women, and important women, were better represented according to the network measures discussed was the one constructed from the \textit{Odyssey}. The \textit{Odyssey} is of Greek origin but was the only Greek narrative ranked highly in this regard. The other two Greek narratives, the \textit{Iliad} and the \textit{Aeneid}, were ranked 15{th} and 7{th} respectively when considering the percentage of women in the networks. The \textit{Aeneid} is a Latin poem of Virgil's which tells us of a Trojan, Aeneas, who escapes after the war and leads to the beginning of Rome~\cite{Aeneid2004}. Though these two networks were not ranked highly when judging the percentages of important women, they were also not ranked in the bottom three narratives. 

Comparing other narratives to networks of a similar origin indicates that some Irish and Icelandic narratives show a good representation of women while others do not. For example, the networks of \textit{TBC} and \textit{Acallam} were ranked highly when judging the presence of important women in the narratives. Two other narratives, \textit{Ulster} and \textit{Lady Gregory}'s text \cite{LadyGregory}, are of Irish origin and do not always rank highly on its female representation.  Only $8.00\%$ of the top $10\%$ of the \textit{Ulster} cycle network are female characters and this narrative is joint 5{th} from the bottom in this regard. It also ranks lowly when judged on the percentage of important women in the network. \textit{Lady Gregory}'s text is ranked 5{th} in this regard but is average in other comparisons. This comparison has shown that some Irish narratives have a good female representation, including important females, and those that do not are still ranked favourably in some aspects. This may suggest that the Irish culture into which these narratives are set  values women of importance. Again, this is highlighted by one of \textit{TBC}'s main characters being a female; Queen Medb.

Some Icelandic narratives show a good percentage of important or well-connected women  compared with the other networks; \textit{G\'{i}sli S\'{u}rsson's Saga} has the highest percentage of women with a degree more than the network's average. \textit{Laxd{\ae}la Saga} is consistently ranked highly when considering the female characters in the network \cite{IcelandS2000}. Other Icelandic narratives include \textit{Nj\'{a}l's Saga} and \textit{Vatnsd{\ae}la's Saga} \cite{IcelandS2000}. These two narratives, like \textit{Lady Gregory}'s text, are ranked in the middle of the narratives for different measures. While not highlighted earlier in this work, \textit{Vatnsd{\ae}la's Saga}, a saga which tells of one family settling in Iceland, ranked highly for some measures. This included the percentage of women overall, although when considering other measures like the percentage of well-connected women it did not. 

The networks of the \textit{Arthurian Romances} and the \textit{Navaho Indian Myths}, which are the only networks of their respective cultures included in table \ref{Tab20gender}, were both ranked highly for different measures. The \textit{Navaho Indian Myths} was one of the narratives with a better percentage of important women and had the highest percentage of both women overall and edges involving women. The \textit{Arthurian Romances} was also ranked highly on its percentage of important women and female edges. 

Some narratives, such as \textit{Beowulf} and \textit{Popol Vuh}, ranked highly in some regards but lowly in others. \textit{Beowulf} deemed no women important by their betweenness centrality and so was ranked last on this measure but, it also had the second highest percentage of well-connected women. \textit{Popol Vuh} did rank highly in regard to well connected characters although it did not feature in the rankings of other measures which specifically focused on females.

\textit{A Song of Ice and Fire}, one of the fictional narratives, did not show high percentages of female characters but it also did not show the lowest percentages either. This may have been unexpected when considering how females are portrayed/treated by Martin although is less so when examining how all characters are treated by Martin. Similarly, \textit{Ossian}, the lone Scottish narrative, was not always ranked within the bottom three of the networks, but it was never ranked highly. However, our last narrative and the other fictional tale, \textit{LotR} was ranked lowly for many of the measures discussed here including the general proportion of female characters and the percentage of those that are deemed important or well-connected.

\section{Conclusions}

Although we oftener read pieces of fiction and mythology that focus on the trials and journey of leading men rather than women, we can see that in the current society an increasing number of works are telling us the stories of women. However, here we endeavoured to investigate how female characters were presented in these pieces of fiction and mythology which so often centre on a male main character, or on multiple male characters.

We used the methodology of network analysis to examine and compare the social networks present in 21 different narratives. We specifically did this with the intention of focusing on the representation of women within the networks. This analysis highlighted which cultures or societies (as depicted within these narratives) were seen to have a good presence of females, especially important females, and which were not. 

Some cultures were represented by more than one narrative within this study and showed that across various texts women were portrayed well. Amongst these narratives there were Greek epics such as the \textit{Odyssey} and the \textit{Aeneid}. Some narratives of Icelandic and Irish origin were seen to contain good percentages of important women while others less so. The latter includes \textit{Lady Gregory} and \textit{Nj\'{a}l's Saga} while \textit{TBC} and \textit{G\'{i}sli S\'{u}rsson's Saga} exemplify the former. Similarly, both the \textit{Arthurian Romances} and the \textit{Navaho Indian Myths}, which were the only narratives of Welsh and Navaho origin respectively, ranked highly when judged on the percentage of important women and female edges in their network but not in other areas investigated here.

Some narratives neither ranked highly nor unfavourably in any measures meaning they did not include great or terrible representations of women. However, there were some narratives such as the fictional \textit{LotR} trilogy which ranked lowly for many measures including the percentage of women overall and the percentage of those that were important. 

Through these comparisons, we see that while no singular narrative or culture has the best level of female representation, we have observed various steps that can be taken to explore the positions of female characters and as such have demonstrated how these network science techniques can be employed to further investigate the social groups depicted within the narratives on an attribute level.

\section*{Acknowledgements}

We wish to acknowledge the efforts of those who helped with the data gathering. This paper is part of a project with Tata Steel whom we also thank for their support.

\bibliographystyle{cmpj}
\bibliography{References}
\nocite{newman2010,kenna2022network,janickyj2022enigmatic,freeman1977set,bonacich1987power,kenna2016maths, kenna2017myths, mac2012universal,yose2018CGG, newman2003social, scott1991, prugl2003gender}

\ukrainianpart

\title{Репрезентація жінок у різних міфологіях}
\author{М. Яніцкі\refaddr{label1,label2,label4}, 
	П. Маккаррон\refaddr{label3},
	Дж. Хосе\refaddr{label1,label2},
	\framebox{Р. Кенна}\refaddr{label1,label2}}
\addresses{
	\addr{label1} Центр дослідження рідинних та складних систем, Університет Ковентрі, Ковентрі, CV1 5FB, Великобританія
	\addr{label2} Колаборація $\mathbb{L}^4$ та докторантський колледж зі статистичної фізики складних систем, Ляйпциг-Лотарингія-Львів-Ковентрі, Європа
	\addr{label3} Факультет математичної статистики, Університет Лімерика, Лімерик V94 T9PX, Ірландія
	\addr{label4} Факультет комп'ютерних наук, Університетський коледж Лондона, Лондон, Великобританія
}

\makeukrtitle

\begin{abstract}
	\tolerance=3000%
Соціальні групи вивчалися впродовж усієї історії, щоб зрозуміти, як різні конфігурації впливають на тих, хто входить до них. Разом з цим, з’явилося певне зацікавлення щодо соціальних груп як художніх, так і міфо\-логiч\-них творів. Протягом останнього десятиліття ці соціальні групи досліджувалися крізь призму наук про мережі, що дозволило порівняти ці історії на новому рівні. Ми використовуємо цей підхід, щоб зосередитися на атрибутах персонажів у цих мережах, зокрема на їхній статі. Таким чином, ми розглядаємо, як зображається жіноче населення у різних наративах і, певною мірою, у суспільствах, в яких воно знаходиться. Завдяки цьому ми виявили, що, незважаючи на відсутність тенденції, що усі наративи одного походження мають однакове представлення, деякі з них є помітно кращими за інші. Ми також спостерігаємо, які наративи загалом надають перевагу важливим жіночим персонажам, а які ні.
	\keywords мережі, складні системи, соціальні системи
	
\end{abstract}

\end{document}